# End-to-End CNN+LSTM Deep Learning Approach for Bearing Fault Diagnosis

*Amin Khorram[a,]\*, Mohammad Khalooei[b], Mansoor Rezghi[c]*

[a]*Dept of Mechanical Engineering, WSE Co.,1416844816 Tehran, Iran.*
[b]*Dept of IT and Computer Engineering, Amirkabir University of Technology,15875-4413Tehran, Iran.*
[c]*Dept of Computer Science, Tarbiat Modares University,14115-111 Tehran, Iran.*

A B S T R A C T

Fault diagnostics and prognostics are important topics both in practice and research. There is an intense pressure on industrial plants to continue reducing unscheduled downtime, performance degradation, and safety hazards, which requires detecting and recovering potential faults in its early stages. Intelligent fault diagnosis is a promising tool due to its ability to rapidly and efficiently processing collected signals and providing accurate diagnosis results. Although many studies have developed machine leaning (M.L) and deep learning (D.L) algorithms for detecting the bearing fault, the results have generally been limited to relatively small train and test datasets and the input data has been manipulated (selective features used) to reach high accuracy. In this work, the raw data, collected from accelerometers (time-domain features) are taken as the input of a novel temporal sequence prediction algorithm to present an end-to-end method for fault detection. We use equivalent temporal sequences as the input of a novel Convolutional Long-Short-Term-Memory Recurrent Neural Network (CRNN) to detect the bearing fault with the highest accuracy in the shortest possible time. The method can reach the highest accuracy in the literature, to the best knowledge of the authors of the present paper, voiding any sort of pre-processing or manipulation of the input data. Effectiveness and feasibility of the fault diagnosis method are validated by applying it to two commonly used benchmark real vibration datasets and comparing the result with the other intelligent fault diagnosis methods.

## 1. Introduction

Bearings are the essential components in rotary machines. The bearing fault is one of the main reasons for motor failure and to detect the fault in primary stages can prevent great down-time and recovery costs [1]. In recent years implementation of M.L or D.L in many scientific fields has been drastically increased. Intelligent fault detection is one of the areas which has received wide attention and used in practical situations. The key issue of applying M.L techniques into bearing fault diagnosis is developing a network architecture that can get satisfactory diagnosis performance in a relatively short time [2]. Mainly, data-driven intelligent fault detection of bearing is conducted using signal processing approaches. These signals are "vibration signal" or "motor current signal" that are measured using accelerometers or frequency inverters [3], respectively. In the literature, the vibration signal has received more attention due to the more accurate results[4]. To implement M.L or D.L techniques for bearing fault detection, we need to extract features and use them in the learning algorithm aiming to reach the highest accuracy. Features can be categorized in three different domains as: time-domain [5], frequency-domain [6], or time-frequency-domain [7]. In the past decade, M.L techniques such as k-nearest-neighbour (KNN), support-vector-machine (SVM), and artificial-neural-network (ANN) had been promising tools for bearing fault detection [2]. However, the output of those methods is typically acceptable in case of relatively small-scaled-data [8]. For instance, in [9], Yaqub et al use KNN for bearing fault diagnosis and test a small data-frame, they also use higher-order-

———————
 \* *Corresponding author.*
 E-mail address: amin.khorram@rockets.utoledo.edu

cumulants (HOC) and wavelet transform (WT) for the pre-determined transformation of data, nevertheless, do not reach an acceptable accuracy. In [10], Hu et al. use SVM, in the same way, the data is pre-processed and the size of the data-frame is relatively small.

In recent years, with the quick development of advanced measurement techniques, massive data are collected and most of the mentioned conventional M.L algorithms have drawbacks to establish decision models on these data[11]. Hence, tendencies have shifted from conventional methods to more complex ones such as the deep neural network (DNN), the convolutional neural network (CNN), the recurrent neural network (RNN), etc. In [12], Eren et al. utilize a one-dimensional convolutional neural network for time series prediction applied to pre-processed data. Particularly, the input data is filtered, decimated, and normalized to reach a better efficiency. In [11], Zhang et al. claim to reach a high accuracy using DNN for time sequence prediction, however, they do not provide any architecture for their proposed DNN network. The same issue happens in [2], where Mao et al. claim to reach a high accuracy using a novel deep learning method. Although, the authors provide just the training accuracy (not the testing accuracy) and do not provide any feasible architecture for their proposed network which prevents further reproducibility. We could also notice some state-of-the-art articles like [13] and [14] which have simultaneously paid attention to CNN and Long-Short-Term-Memory (LSTM) networks for bearing fault diagnosis. However, the architecture and the step-by-step path they undergo to reach their proposed model are not clearly explained.

All the former efforts in bearing fault diagnosis have the following shortcomings: 1. The features are manipulated or selected. 2. The scale of the dataset is relatively small and cannot cover the comprehensive data on an industrial scale. 3. The accuracies are relatively high but not enough for counting on the outcome in industrial scale. 4. The neural network's architectures are barely presented and the path to reach the claimed accuracy is not evident.

In this work we are going to use a CNN-LSTM network for temporal sequence prediction of the data obtained in the time domain to reach the highest accuracy in a relatively short time. Compared to the other articles in the literature, we are not doing any pre-processing or manipulation of the raw data. As a result, the model can be utilized in a practical situation and extract the real characteristic of the practical system's signal under all circumstances. We have evaluated our proposed model by testing two benchmark bearing datasets in the literature: Intelligent Maintenance Systems (IMS) bearing dataset [15] which is a run to failure raw bearing dataset measured by Centre of Intelligent Maintenance Systems of University of Cincinnati, and the Case Western Reserve University(CWRU) bearing dataset[16] measured by the Bearing Data Centre of the Case Western Reserve University. The result shows that the average accuracy rate in the train and test datasets of the proposed method outreaches the state-of-the-art articles in a relatively shorter interval. Moreover, we are going to provide the architecture and the step by step path we undergo to reach the high accuracy for our proposed fault detection methodology.

The contribution of the following paper can be summarized as follows:

1-Using a relatively bigger data-set for training and testing, compared to the other papers in the literature and achieving a higher generalization accuracy at the shortest possible time (relatively shorter time and smaller number of epochs compared to the articles in the literature).

2-A novel deep learning structure is proposed for bearing fault diagnosis which is highly resistant to overfitting. By applying the proposed deep learning method, this paper can effectively utilize the time series to improve the diagnosis accuracy and the numerical stability for the bearing fault.

3-The model is end-to-end and can be fed by the raw vibration data directly. As a result, no pre-processing such as pre-determined transformation (such as Fast Fourier Transform (FFT) or Discrete Wavelet Transform (DWT)), manipulated feature extraction, and feature selection is required.

The paper is organized as follows. In section 2, a brief review of CNN-LSTM and the network architecture used in this work is presented. In section 3, we describe the test rigs, the datasets, the fault classification and the experiments behind selecting hyper-parameters of our proposed CRNN architecture. Section 4 is devoted to discussion and comparison of our method with the other methods in the literature and finally, section 5 represents the conclusion and the future works.

## 2. Method

### 2.1 CNN+LSTM network

CNNs are biologically inspired feed-forward ANNs which are considered as simple computational models of the mammalian visual cortex[17]. 2D-CNNS and 3D-CNNs are generally used for image and video processing, while 1D-CNNs are mainly used for audio and text recognition (as a time series data). 1D-CNNs are perfect tools for time-series recognition and prediction. The network has recently been used in state-of-the-art applications such as early diagnosis, structural health monitoring, anomaly detection and identification[18]. Considering that our data is of vibration signal (time-series type); we are going to use 1D-CNNs. In this network, the output of a convolutional layer ( $v_{ij}^x$ ) at position x of the $j_{th}$ feature map in the $i_{th}$ layer is denoted as follows[19]:

$$v_{ij}^x = g\left(b_{ij} + \sum_m \sum_{p=0}^{P_i-1} w_{im}^p v_{(i-1)m}^{x-p}\right) \tag{1}$$

Where m indexes the feature map in the previous layer ($(i-1)_{th}$ layer) connected to the current feature map; $w_{im}^p$ denotes the weight of position p in the $m_{th}$ feature map; $P_i$ is the width of the kernel toward the spectral dimension; $b_{ij}$ is the bias of $j_{th}$ feature map in the $i_{th}$ layer and g is the activation function. Regularly, after one or more CNN layers, a Pooling layer is used which can offer invariance by reducing the resolution of the feature maps[20]. Each pooling layer corresponds to the previous convolutional layer. The most common pooling operation is the max-pooling:

$$\overline{u}_n = \max_{1 \le j \le k}(u_n^j) \tag{2}$$

Where $u_n^j$ is the $j_{th}$ element of the $n_{th}$ patch, $\overline{u}_n$ is the sample of the $n_{th}$ patch built by max-pooling and "k" is the size of the $n_{th}$ patch.

LSTM networks are recurrent neural networks equipped with a special gating mechanism that controls access to memory cells [21]. Since the gates can prevent the rest of the network from modifying the contents of the memory cells for multiple time steps, LSTM networks preserve signals and propagate errors for much longer than ordinary recurrent neural networks. LSTM was designed by Hochreiter et al. [21] in order to model temporal sequences and their long-range dependencies more accurate than conventional RNNs. Each LSTM block consists of three gates, namely the input gate, the forget gate and the output gate. Gates are a way to optionally let the information through; they are composed out of a sigmoid activation function which outputs numbers between zero and one:

$$\sigma(t) = \frac{1}{1 + e^{-t}} \tag{3}$$

The value of zero means nothing passes through and the value of one means everything passes through the gate. The equations for the gates are as the following[22]:

$$i_t = \sigma(w_i[h_{t-1}, x_t] + b_i) \tag{4}$$

$$f_t = \sigma(w_f[h_{t-1}, x_t] + b_f) \tag{5}$$

$$o_t = \sigma(w_o[h_{t-1}, x_t] + b_o) \tag{6}$$

Where $i_t$ is the input gate, $f_t$ is the forget gate, $o_t$ is the output gate, $\sigma$ is the sigmoid activation function, $w_x$ is the weight of the respective gate (x), $h_{t-1}$ is the output of the previous LSTM block at time-step t-1, $x_t$ is the input at the current time-step and $b_x$ is the biases for the respective gate (x).
The cell state vector ($c_t$) and output vector of LSTM unit ($h_t$) can be calculated as the following:

$$\tilde{c}_t = \tanh(w_c[h_{t-1}, x_t] + b_c) \tag{7}$$

$$c_t = f_t * c_{t-1} + i_t * \tilde{c}_t \tag{8}$$

$$h_t = o_t * \tanh(c_t) \tag{9}$$

Where $\tilde{c}_t$ represents the candidate vector for cell state vector (how much we decide to update each state vector); tanh is the hyperbolic tangent activation function and (∗) denotes the Hadamard product. In the following, a LSTM block is illustrated:

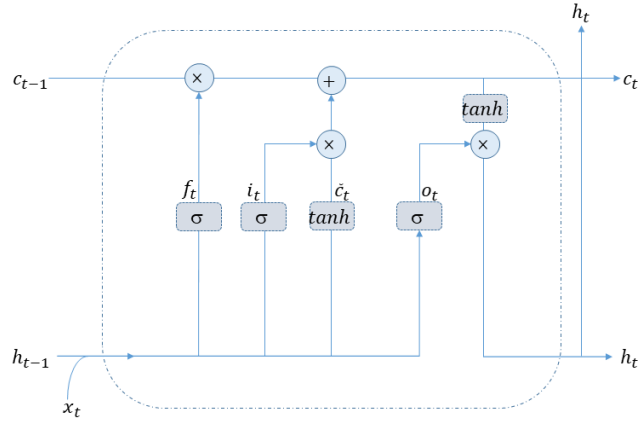

**Fig.1.** LSTM block.

CNN-LSTMs are used for many visual learning tasks but are also known to be used for speech recognition and natural language processing[23]. Moreover, CNNs and LSTM are both powerful tools for temporal sequence prediction[24]. Handling big data or complex temporal sequence problems, CNN-LSTM network enhances accuracy and precision of predictions[25].

To explain the temporal sequence prediction, suppose we observe a dynamical system over a temporal region represented by an M×N grid which consists of M rows and N columns. Inside each cell in the grid, there are P measurements which vary over time. Thus, the number of features can be represented by a tensor of size PxMxN. If the features are recorded periodically, the dataset can be divided into samples of equal temporal length and as a result we have a sequence of tensors $\hat{X}_1, \hat{X}_2, \hat{X}_3, \ldots, \hat{X}_n$. The temporal sequence prediction problem is to predict the most likely $k_{th}$ sequence of the observation, given the previous $j_{th}$ observation by maximizing the following conditional probability[26]:

$$\hat{Y}_{t+1}, \ldots, \hat{Y}_{t+k} = \underset{X_{t+1}, \ldots, X_{t+k}}{\arg\max} \, p(X_{t+1}, \ldots, X_{t+k} \mid \hat{X}_{t-j+1}, \hat{X}_{t-j+2}, \ldots, \hat{X}_t) \tag{10}$$

### 2.2. Architecture and Learning method

As mentioned in the previous section, our data is recorded in the time domain and can be categorized as time-series. Time-series have local and global features and for a model to be highly accurate in processing time-series, it must consider both features at the same time. Time-series have a strong 1D structure: variables (or pixels) that are temporally nearby are highly correlated. Local correlations are the reasons for the well-known advantages of extracting and combining local features before recognizing global features [27]. Convolutional networks force the extraction of local features by restricting the receptive fields of hidden units to be local [27]. On the other hand, LSTMs can learn long-term dependencies between two entities [28] and as a result can handle global features. Consequently, the combination of these two networks allows us to handle the scrutiny of the mentioned data. It is worth mentioning that, CNNs in general, have a de-noising property that could reduce the effect of the noise in the learning process (LSTMs on the contrary are sensitive to noise) and require relatively little pre-processing compared to the other temporal prediction methods[29]. Moreover, the CNN-LSTM model is more efficient in preventing overfitting compared to the other DNN models. Considering the mentioned advantages and the numerous experiments fulfilled to find the best model that reaches the highest accuracy at the shortest possible interval, an improved CNN-LSTM model is chosen for bearing fault diagnosis. It must be noted that, if the CNNs and LSTMs are displaced (the input enters to LSTM network and CNNs afterward) not only the noise will deteriorate the result but also the data will be processed globally in the first place and the local feature extraction remains inefficient.

The raw vibration datasets or features are collected in the time domain. The following figure represents the data structure:

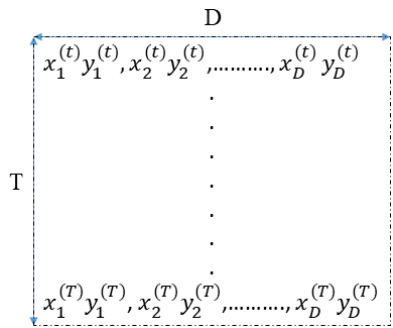

**Fig.2.** Dataset

Where $X_i^{(t)}, Y_i^{(t)}$ represent the variables of $i_{th}$ bearing at time t, D is the number of bearings (test-cases) and T is the total test-time.

Our proposed CNN-LSTM architecture is depicted in Fig.4. The input of our CRNN network are tensors of equal size, therefore, the first step is to divide the dataset into samples or temporal sequences of equal length in order to feed them to our CRNN model. In the next step, the features are split into train, validation, and test sets. The hyper-parameters of the model are obtained by minimizing the cost function (average of loss functions of the entire training set). As it can be observed in Fig.4, the proposed architecture consists of 1D-CNNs and LSTMs layers. The proposed network is tested in section 3 using a big dataset (IMs bearing dataset) in order to find the optimum hyper-parameters of which minimize the cost function efficiently (the smaller datasets such as the one we select from CWRU bearing dataset, present more accurate results due to slighter risk of overfitting). Throughout our experiments, an acceptable accuracy is obtained using a 1D-CNN of 84x84 dimensionality and an LSTM containing 24 neurons. There is one dropout layer after each main layer, a dropout layer effectively prevents overfitting by reducing the correlation between neurons[30].

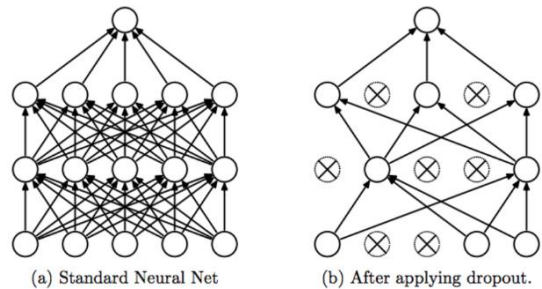

(a) Standard Neural Net                    (b) After applying dropout.

**Fig.3.** Dropout Neural Net Model. Left: A standard neural net with 2 hidden layers. Right: An example of a thinned net produced by applying dropout to the network on the left. Crossed units have been dropped [30].

We have also used batch-normalization layers in order to speed up and enhance the stability of network and the accuracy of learning[31]. Batch-normalization makes networks robust to bad initialization of weights; reduces covariance shift by normalizing and scaling inputs and scale and shift parameters to avoid losing stability of the network [31]. The following equations represent a Batch-Normalization transform applied to activation x over a mini-batch[31]:

$$\text{Input}: \text{mb} = \{x_{1........m}\} \tag{11}$$

$$\text{mean}_{mb} = \frac{1}{n}\sum_{i=1}^{n} x_i \tag{12}$$

$$\sigma_{mb}^2 = \frac{1}{n}\sum_{i=1}^{n}(x_i - \text{mean}_{mb})^2 \tag{13}$$

$$\hat{x}_i = \frac{x_i - \text{mean}_{mb}}{\sqrt{\sigma_{mb}^2 + \varepsilon}} \tag{14}$$

$$\text{Output}: y_i = \gamma\hat{x}_i + \beta = \text{BN}_{\gamma,\beta}(x_i) \tag{15}$$

Where the input is a mini-batch of size m; $\gamma$ and $\beta$ are the parameters to be learned; $\sigma_{mb}^2$ is the mini-batch variance; $\hat{x}_{1,...,m}$ are the normalized values. Finally, the fully-connected layer takes advantage of sigmoid activation. To solve the optimization problem, the Adagrad method is used. In addition, the loss functions used for compiling is mean-squared-logarithmic-error (MSLE). The proposed CRNN network predicts $\hat{y}$, and using the following MSLE loss function, the deviation and accuracy of the model are measured:

$$L(y, \hat{y}) = \frac{1}{N}\sum_{i=0}^{N}(\log(y_i + 1) - \log(\hat{y}_i + 1))^2 \tag{16}$$

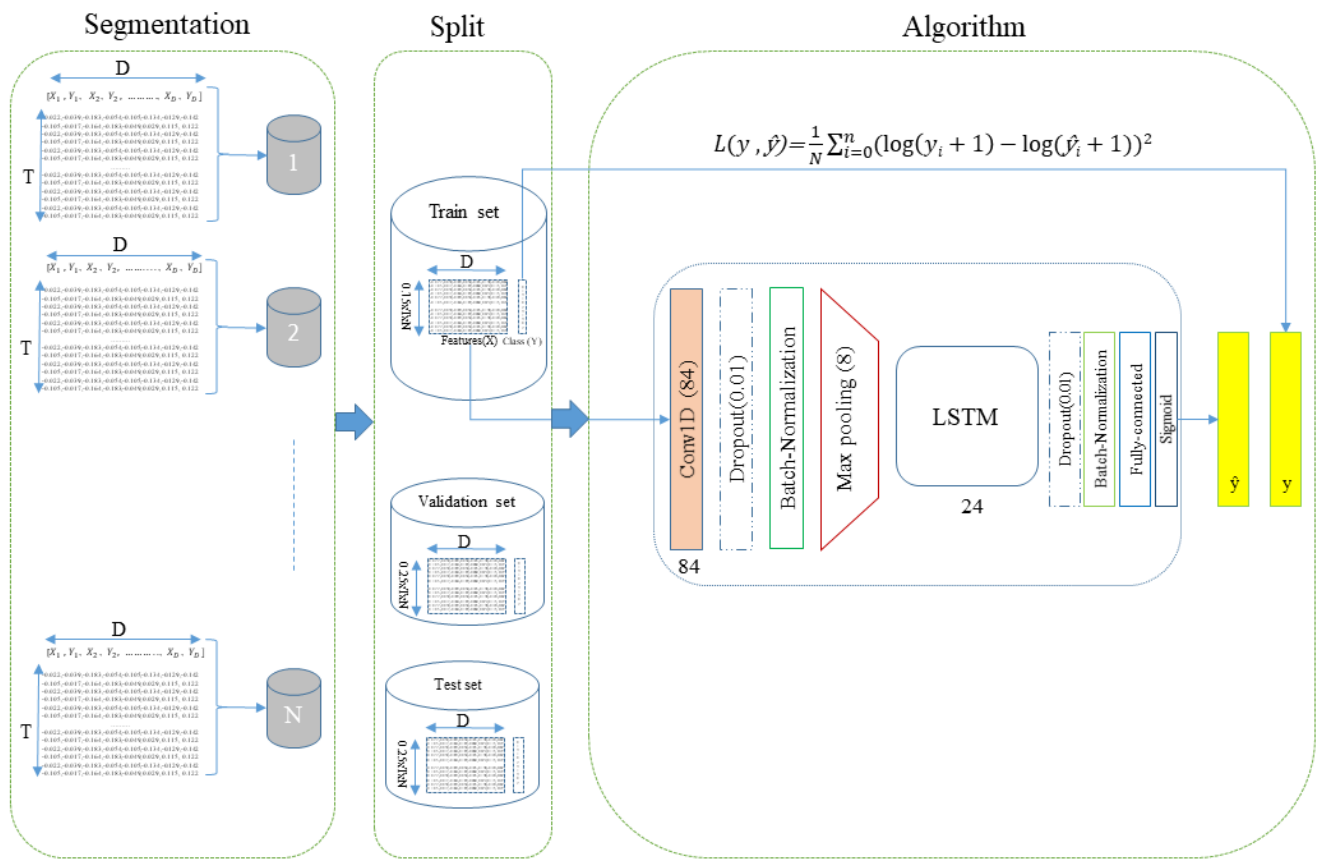

**Fig.4.** The proposed CRNN architecture

## 3. Experiments

In this section, we are going to evaluate our proposed method, by testing it on two benchmark bearing datasets: IMS and CWRU. In the following, we are going to describe the test rigs used for capturing the vibration data, fault classification (labels), and the raw vibration signals (features) which are used as the input of our CRNN algorithm. The section proceeds toward analysing the amplitude/time diagrams of the datasets, the procedure of designing our architecture, and finally, the accuracy/loss diagrams and confusion matrixes for each test.

### 3.1. IMS Bearing Dataset

To validate the proposed method, experimental datasets are applied to test its performance. The first dataset is provided by the University of Cincinnati Center of Intelligent Maintenance Systems [8]. The experimental apparatus is shown in Fig.5.

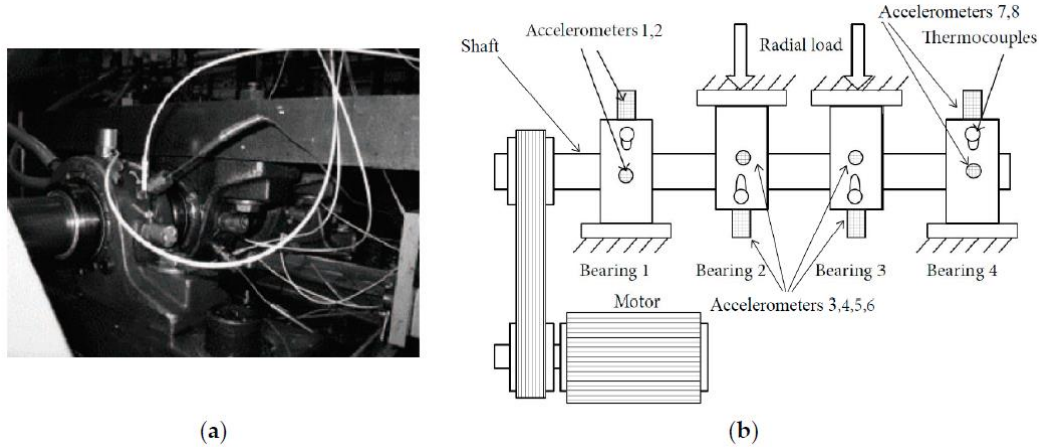

(a)                                                                (b)

**Fig.5.** (a) Image of a bearing with the connected accelerometers. (b) Schematic of the test rig with details.

As depicted in Fig.5, there is a shaft on which four bearings are installed. The bearings model is Rexnord ZA-2115 double-row. Two high precision accelerometers are connected to each bearing in Cartesian coordinates; therefore, the vibration is measured in X and Y directions for each bearing. The shaft is driven by an alternative current (AC) motor which is connected to the shaft using a conveyor belt. The shaft and bearings are under a radial load of 2721.5 Kg imposed by a spring mechanism. The rotation speed of the shaft is 2000 revolution per minute (RPM). The sampling rate is set as 20 KHz and every 20480 data points (recorded in one second) are recorded in a single file. In every 5 or 10 minutes, the data is recorded and written in files while the bearings ae rotating. Each test consists of 2156 files. Therefore, the total number of data-points is 44,154,880 for each test. Previous works done on the 1st-test of IMS bearing dataset shows that there are seven different states of health during the test[32]:

- Early (initial run-in of the bearings)
- Normal
- Suspect (the health seems to be deteriorating)
- Imminent failure (for bearings 1 and 2, which didn't fail, but were prone to damage)
- Inner race failure (bearing 3)
- Rolling element failure (bearing 4)
- Stage 2 failure (bearing 4)

The vibration signals of some states are pretty close which cannot be distinguished by signal processing, therefore to reduce the calculation complexity and improve the performance of our learning algorithm, we choose the labels with the highest importance both in fault detection and practical situation (the other states underlie the following ones):

- Healthy (data taken from early and normal states)
- Suspected
- Inner Race failure
- Rolling element failure

As discussed above, the number of data-points in the 1st-test is pretty big and it is so time and memory consuming to use this big dataset as the input of our learning algorithm. Therefore, we randomly choose 30 files for each class or state of health. In the next step, data is concatenated, labelled and prepared to be fed to the learning algorithm. Labels are 0-(Healthy), 1-(Suspected), 2-(Inner-race-fault) and 3-(Rolling-element-fault). As explained in the previous section we are using a CRNN network; hence, the input is supposed to be sequence of tensors with equal dimensions. The sampling rate is 20KHz and the rotation speed is 2000RPM, so it can be calculated that there are 600 points per revolution (rotation period). The size of each sample is set to be a quarter of the rotation period, which is 150 rows of data. In each row of data, we have the vibration data of bearings in X and Y directions. Therefore, there are 8 features in each row namely:

- $X_1$ , $Y_1$ (measured by accelerometers connected on the first bearing)

- X₂ , Y₂(measured by accelerometers connected on the second bearing)
- X₃ , Y₃(measured by accelerometers connected on the third bearing)
- X₄ , Y₄ (measured by accelerometers connected on the fourth bearing)

As a result, each sample is a tensor of (150x8x1) dimension and the input tensor for each health state has a dimension of (4096x150x8). The total number of samples is 16384 with 4096 samples for each health state. The number of samples for each class can be observed in Table 1. The Amplitude/time diagram of the four health states is depicted in Fig.6. As it can be observed, each health state has a specific vibration signal signature.

**Table 1 – Number of samples and class number for each health state, IMS dataset.**

| State | Number of samples | Class (Label) |
|---|---|---|
| Healthy | 4096 | 0 |
| Suspected | 4096 | 1 |
| Inner-race-fault | 4096 | 2 |
| Rolling-element-fault | 4096 | 3 |

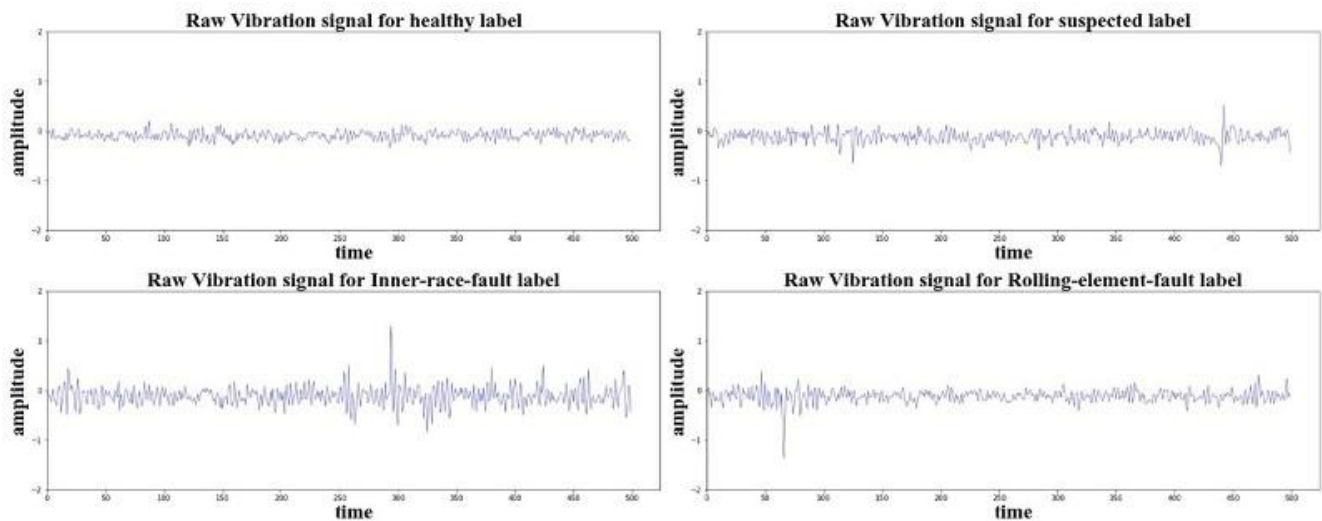

**Fig.6.** IMS Bearing Dataset, raw vibration signal for Healthy, Suspected, Inner-race-fault and Rolling-element-fault.

In the next step, data is split into train, validation and test sets. To take advantage of a stateful LSTM network, the split and batch-size selection should be performed in a way so that the number of samples in train and test sets be integers and divisible by the batch-size. The number of samples for all the four classes are 16,384. Therefore, we allocate 25% of dataset to the train and 75% to the test sets and the optimum batch-size turns to be 64.

To select the best architecture for our model, different networks with various hyper-parameters are experimented. The number of epochs for all the accomplished tests is set to 50 and the goal is to reach the highest training accuracy at the shortest possible time. The simulation model is based on the Tensorflow library in the Python. The Processor is Intel(R) Core(TM) i7-8550U CPU @ 1.80GHz, 1992 MHz, 4 Core(s), 8 Logical Processor(s) and Physical Memory (RAM) is 8GB.

The following tables represent part of the numerous experiments we fulfilled in order to find the optimum hyper-parameters for our proposed model:

**Table 2 –Finding the optimum hyper-parameters for the represented model.**

| Test No. | Conv1D Filters | Conv1D Kernel-size | Convolutional Activation | First Dropout | Maxpooling Size | LSTM Neurons | Second Dropout | Keras Loss | Keras Optimizer | Fully-Connected-Activation | Train Accuracy | Test Accuracy | Calculation Time(seconds) |
|---|---|---|---|---|---|---|---|---|---|---|---|---|---|
| 1 | 16 | 4 | elu | 0.01 | 8 | 24 | 0.01 | MSLE | Adagrad | sigmoid | 0.9284 | 0.9087 | 59 |
| 2 | 16 | 8 | elu | 0.01 | 4 | 24 | 0.01 | MSLE | Adagrad | sigmoid | 0.9587 | 0.9250 | 90 |
| 3 | 32 | 8 | elu | 0.01 | 8 | 12 | 0.01 | MSLE | Adagrad | Sigmoid | 0.9721 | 0.9138 | 74 |
| 4 | 32 | 8 | elu | 0.01 | 8 | 24 | 0.01 | MSLE | Adagrad | Sigmoid | 0.9772 | 0.9294 | 92 |
| 5 | 32 | 8 | elu | 0.01 | 8 | 36 | 0.01 | MSLE | Adagrad | Sigmoid | 0.9734 | 0.9224 | 100 |
| 6 | 32 | 32 | relu | 0.01 | 8 | 24 | 0.01 | MSLE | RMSprop | Sigmoid | 0.9983 | 0.9517 | 149 |
| 7 | 64 | 32 | elu | 0.01 | 8 | 12 | 0.01 | MSLE | Adagrad | Sigmoid | 0.9992 | 0.9502 | 244 |

| Test No. | Conv1D Filters | Conv1D Kernel-size | Convolutional Activation | First Dropout | Maxpooling Size | LSTM Neurons | Second Dropout | Keras Loss | Keras Optimizer | Fully-Connected Activation | Train Accuracy | Test Accuracy | Calculation Time(seconds) |
|---|---|---|---|---|---|---|---|---|---|---|---|---|---|
| 8 | 64 | 32 | elu | 0.01 | 8 | 24 | 0.01 | MSLE | Adagrad | Sigmoid | 0.9977 | 0.9521 | 269 |
| 9 | 64 | 64 | elu | 0.01 | 8 | 24 | 0.01 | MSLE | Adagrad | Sigmoid | 0.9998 | 0.9641 | 484 |
| 10 | 84 | 64 | elu | 0.01 | 8 | 24 | 0.01 | MSLE | Adagrad | Sigmoid | 0.9999 | 0.9595 | 334 |
| 11 | 84 | 84 | elu | 0.01 | 8 | 12 | 0.01 | MSLE | Adagrad | Sigmoid | 1.000 | 0.9619 | 538 |
| **12** | **84** | **84** | **elu** | **0.01** | **8** | **24** | **0.01** | **MSLE** | **Adagrad** | **Sigmoid** | **1.000** | **0.9713** | **419** |
| 13 | 96 | 96 | elu | 0.01 | 8 | 24 | 0.01 | MSLE | Adagrad | Sigmoid | 0.9999 | 0.9666 | 680 |
| 14 | 128 | 84 | elu | 0.01 | 8 | 24 | 0.01 | MSLE | Adagrad | Sigmoid | 1.000 | 0.9639 | 661 |

The best result is achieved in test no.12. As can be inferred, increasing the kernel size results in a more generalized snapshot of the input [33]. As can be inferred, the accuracy of the model improves dramatically by increasing the Conv1D hyper-parameters up to a certain level. The enhancement acts diversely from that point on. Tests no.13 and no.14 reveal that exaggerating the number of filters or kernel size for the Conv1D layer, not only deteriorates the test accuracy but also increases the calculation time dramatically. The best value for the LSTM neurons is obtained in test no.4. In this test, the number of LSTM neurons is modified to 24 and the accuracies are improved clearly. Test no.5 expresses that further enhancement of LSTM neurons does not improve the network performance.

In the following table, some multi-layer networks are evaluated to clarify the effect of additional layers on the accuracy:

**Table 3 –Effect of additional layers on the accuracy.**

| Test No. | Conv1D Filters | Conv1D Kernel-size | Convolutional Activation | First Dropout | Maxpooling Size | LSTM Neurons | Second Dropout | Keras Loss | Keras Optimizer | Fully-Connected Activation | Train Accuracy | Test Accuracy | Calculation Time(seconds) |
|---|---|---|---|---|---|---|---|---|---|---|---|---|---|
| 1 | $\frac{32}{16}$ | $\frac{8}{4}$ | elu | 0.01 | 8 | 24 | 0.01 | MSLE | Adagrad | Sigmoid | 0.9735 | 0.9248 | 101 |
| 2 | $\frac{32}{16}$ | $\frac{8}{4}$ | elu | 0.01 | 8 | $\frac{24}{12}$ | 0.01 | MSLE | Adagrad | Sigmoid | 0.9803 | 0.9172 | 98 |
| 3 | 64 | 32 | elu | 0.01 | 8 | $\frac{24}{12}$ | 0.01 | MSLE | Adagrad | Sigmoid | 0.9998 | 0.9529 | 538 |
| 4 | 84 | 84 | elu | 0.01 | 8 | $\frac{24}{12}$ | 0.01 | MSLE | Adagrad | Sigmoid | 0.9999 | 0.9675 | 500 |
| 5 | $\frac{84}{16}$ | $\frac{84}{4}$ | elu | 0.01 | 8 | 24 | 0.01 | MSLE | Adagrad | Sigmoid | 0.9989 | 0.9614 | 536 |

In Table.3, test no.1 consists of two Conv1D layers with 32 and 16 number of filters, and kernel-size of 8 and 4 respectively. Comparing the result with test no.4, in Table.2 we can conclude that increasing the number of layers has a negative effect on the accuracy. In the following, additional layers are added to the optimum network (Test no.12 in Table 2).  Comparing the accuracy of test no.4 and no.5 in Table.3 with test no.12 in Table.2, the negative effect of additional layers is evident.

Finally, in the following table, the rest of hyper-parameters (Activation functions, Loss functions, etc.) for the optimum network are evaluated:

**Table 4 –Evaluating the rest of hyper-parameters.**

| Test No. | Conv1D Filters | Conv1D Kernel-size | Convolutional Activation | First Dropout | Maxpooling Size | LSTM Neurons | Second Dropout | Keras Loss | Keras Optimizer | Fully-Connected Activation | Train Accuracy | Test Accuracy | Calculation Time(seconds) |
|---|---|---|---|---|---|---|---|---|---|---|---|---|---|
| 18 | 84 | 84 | relu | 0.01 | 8 | 24 | 0.01 | MSLE | RMSprop | Sigmoid | 0.9993 | 0.9563 | 601 |
| 19 | 84 | 84 | relu | 0.01 | 8 | 24 | 0.01 | MSLE | Adagrad | relu | 0.9994 | 0.9456 | 597 |
| 20 | 84 | 84 | sigmoid | 0.01 | 8 | 24 | 0.01 | MSLE | Adagrad | Sigmoid | 1.000 | 0.9561 | 507 |
| 21 | 84 | 84 | elu | 0.01 | 8 | 24 | 0.01 | MSLE | Adam | Sigmoid | 0.9999 | 0.9668 | 626 |
| 22 | 84 | 84 | elu | 0.01 | 8 | 24 | 0.01 | Cat-Cr-entropy | Adagrad | Sigmoid | 1.000 | 0.9641 | 539 |

Throughout the numerous experiments we fulfilled (of which only some are represented in Tables.2, Table.3 and Table.4) the most acceptable hyper-parameters for our proposed model are embolden in Table.2, test no.12. which contains a conv1D layer of 84 filters with kernel size of 84, plus a LSTM layer containing 24 neurons. The best activation for Conv1D and Fully-connected layers are elu and sigmoid respectively and the best keras loss/optimizer functions are MSLE and Adagrad respectively. The best train and test accuracies are 1.000 and 0.9713 respectively and the computation time of the test iss 419 seconds. The schematic of the optimum architecture is presented in Fig.4.
The train and test accuracies/losses diagrams of the optimum result on IMS dataset can be observed in Fig.7, and the confusion matrix for this test is depicted in Fig.8.

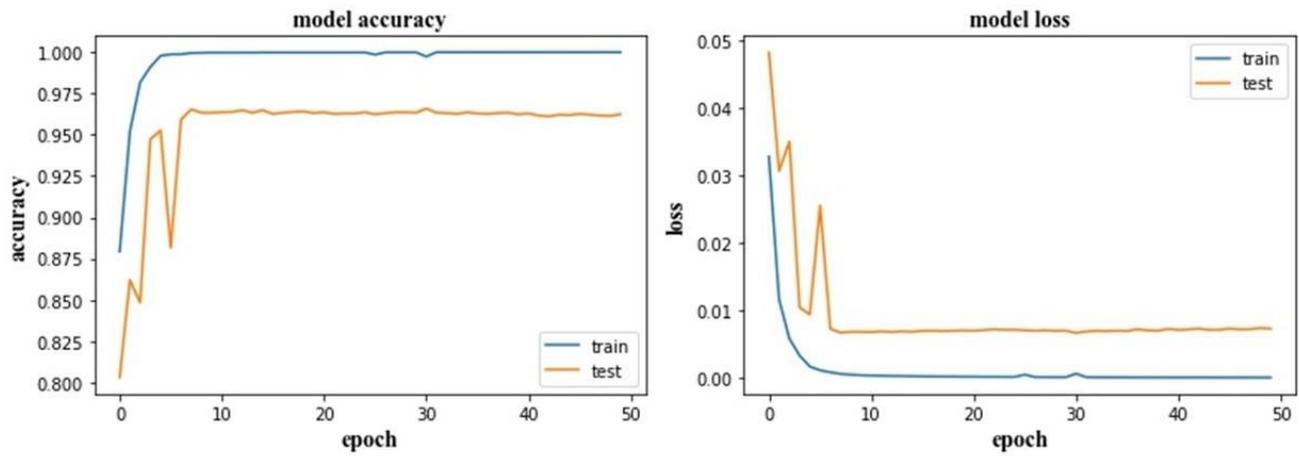

**Fig.7.** IMS Bearing Dataset, train/test accuracy and loss diagrams.

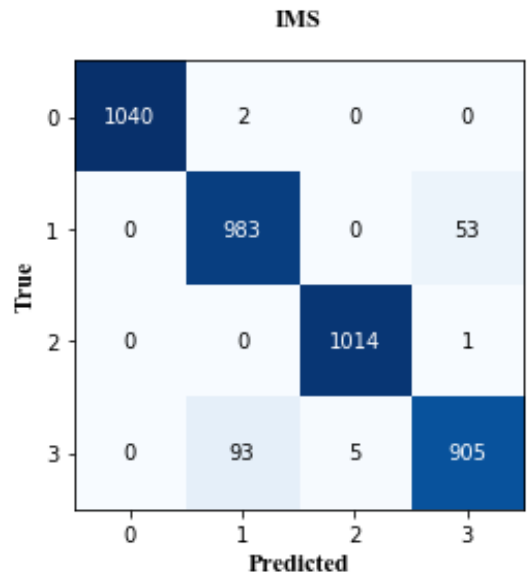

**Fig.8.** Confusion matrix of the test on the IMS dataset.

As it can be observed in Fig.8, the classifier has missed some predictions in classes 1 and 3 or the "suspected state" and "outer-race-fault state". Returning to Fig.6, the signal diagram of the two classes are pretty close and confusing for the CRNN model. However, considering that we have not used any data pre-processing, data selection or manipulation, the strength of the model in Health vs Fault state diagnosis for this test is acceptable.

We have reached a high fault detection accuracy for IMS bearing dataset, implementing our CRNN algorithm. In the next step we are going to test our proposed model for the second benchmark bearing dataset.

### 3.2 CWRU Bearing Dataset

The CWRU dataset is provided by Case Western Reserve University Bearing Data-Center[16]. The test rig is shown in Fig.9.

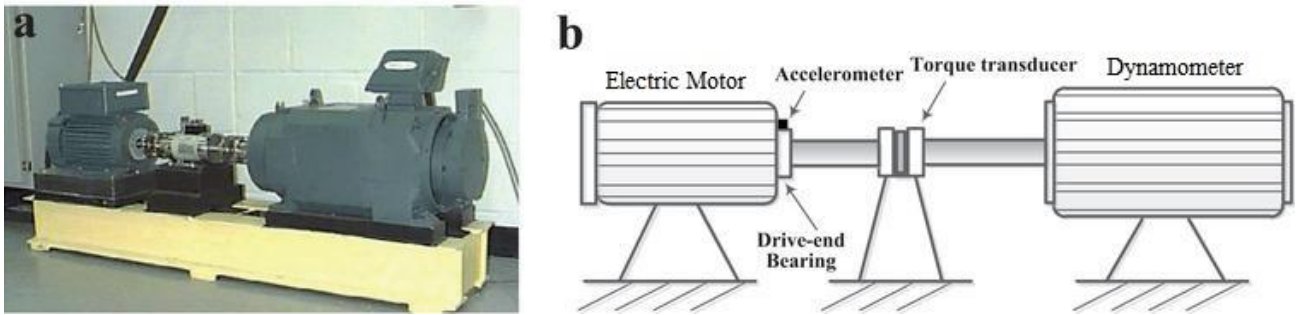

**Fig.9.** (a) Image of the test rig (b) Schematic of the test rig with details.

The test rig consists of a 2-hp Reliance Electric motor, bearing and fastened accelerometer, a torque transducer/encoder and a dynamometer. The tested bearing is SKF deep-groove ball bearings 6205-2RS JEM. Accelerometer is placed at the 12 o'clock position of the motor housing. Data is collected at 12 KHz for drive-end-bearing-experiment. Single point fault is introduced to the test bearing using electro-discharge machining with fault diameters of 0.53$_{mm}$ at the inner raceway, rolling element and outer raceway. The approximate motor speed is 1750 rpm. There are 6 states of health for this test:

- Normal (Healthy)
- Ball Fault
- Inner-Race-Fault
- Outer-Race-Fault at 3 o'clock (Fault placed at the load zone)
- Outer-Race-Fault at 6 o'clock (Fault placed orthogonal to the load zone)
- Outer-Race-Fault at 12 o'clock (Fault placed orthogonal to the load zone)

We choose 121155 data-points for each state of health. Implementing our CRNN model, the input is supposed to be sequence of tensors with equal dimensions. Given the sampling rate of 12KHz and rotation speed of 1750rpm, there are approximately 411 points per revolution (rotation period). We choose the number of data-points in each sample to be 205 corresponding to half a revolution approximately, to reach the highest train and test accuracy. The number of data-points for each health state is also selected so that it would be divisible by the number of data-point per sample. Therefore, each sample is a tensor of (205x1x1) dimension and the input tensor for each health state has the dimension of (591x205x1).

The Amplitude/time diagram of the six health states is illustrated in Fig.10. As can be observed, each health state has a specific vibration signal signature.

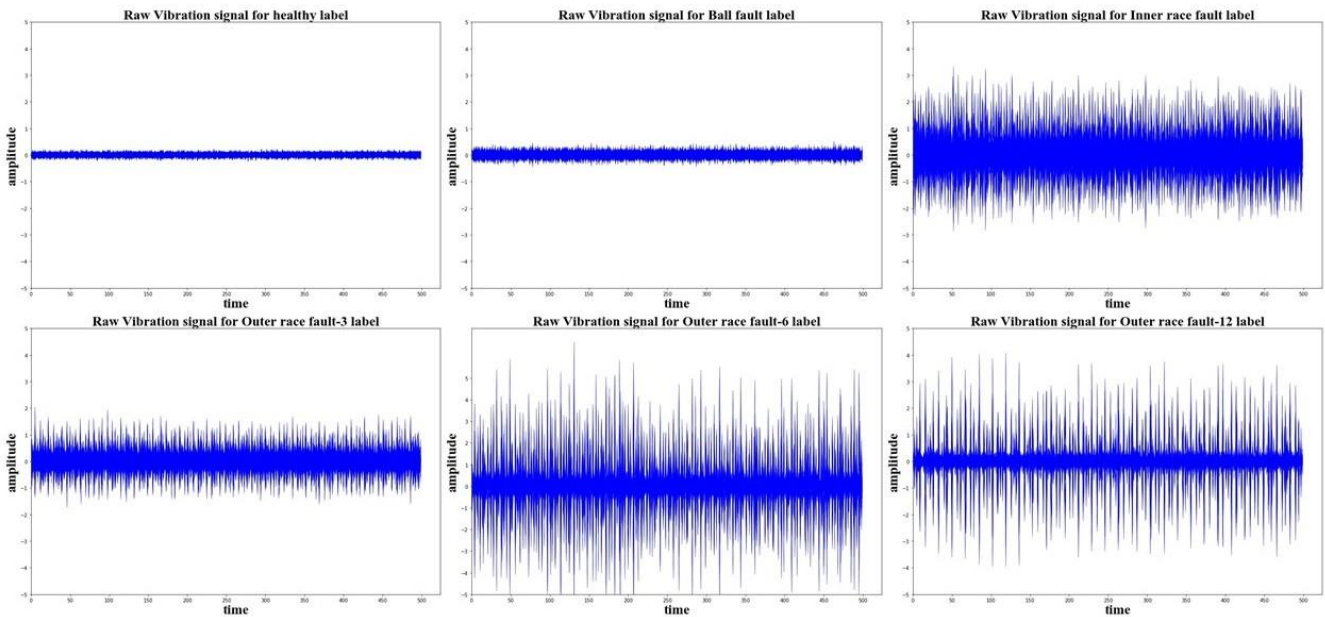

**Fig.10.** CWRU Bearing Dataset, raw vibration signal for different states.

**Table 5 – Number of samples and class number for each health state, CWRU dataset.**

| State | Number of Samples | Class(Label) |
|---|---|---|
| Healthy | 591 | 0 |
| Ball Fault | 591 | 1 |
| Inner-Race-Fault | 591 | 2 |
| Outer-Race-Fault-3 o'clock | 591 | 3 |
| Outer-Race-Fault-6 o'clock | 591 | 4 |
| Outer-Race-Fault-12 o'clock | 591 | 5 |

The samples of equal size are split into train, validation and test sets. To take advantage of a stateful LSTM network, the split and batch size selection should be performed in a way so that the number of samples in train and test sets be integers and divisible by the batch size. The number of samples for all the six classes is 3546, therefore, we allocate 50% of the dataset to train and 50% to test and the optimum batch-size would be 197. Consequently, the accuracies of train and test for 50 epochs turn to be 1.0000 and 0.9977 respectively. The calculation time for 50 epochs is 61 seconds.

The train and test accuracies/losses diagrams of the test on the CWRU dataset can be observed in Fig.11, and the confusion matrix for the test is depicted in Fig.12.

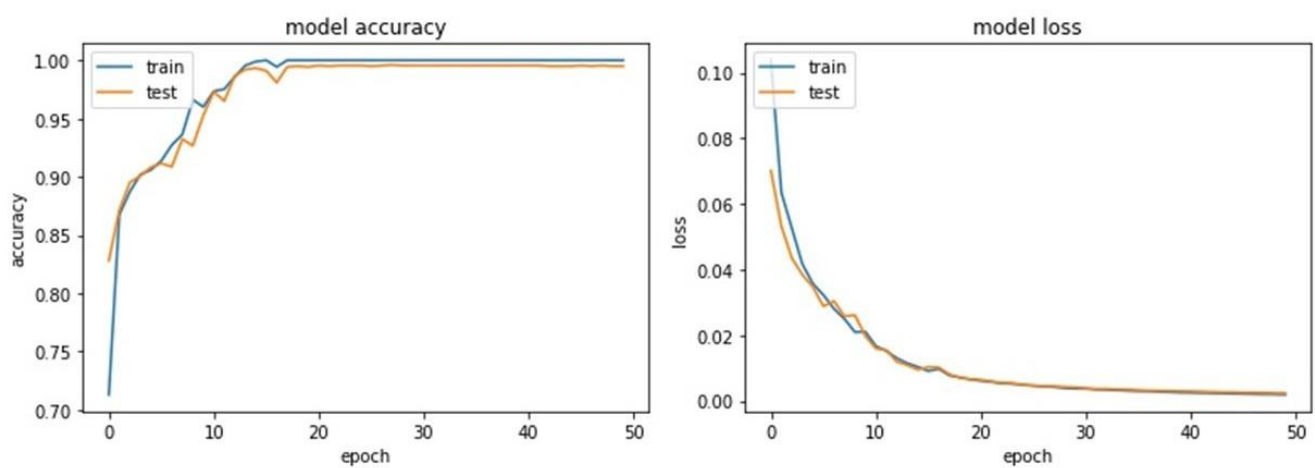

**Fig.11.** CWRU bearing dataset, train/test accuracy and loss diagram.

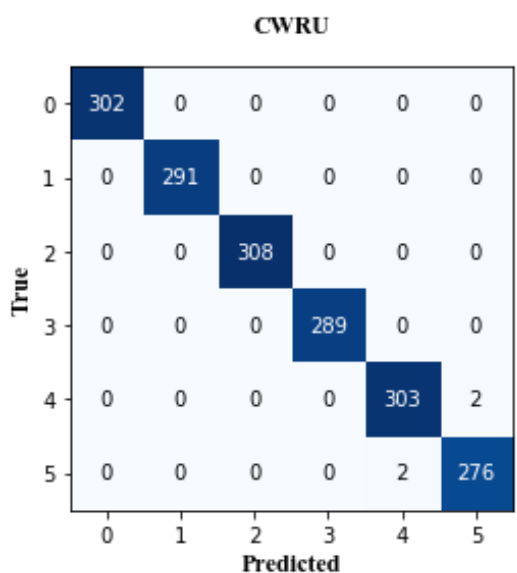

**Fig.12.** Confusion matrix of the test on the CWRU dataset.

As can be observed in Fig.12, the classifier has almost predicted all the six classes correctly. It only missed some predictions in classes number 4 and 5 or the "outer race fault-6 o'clock" and "outer race fault-12 o'clock" labels. Although returning back to Fig.10, the signal diagrams of the two classes are pretty close, and considering that both faults are on the outer race, the error can be overlooked. Taking into consideration that we have not used any data pre-processing, data selection or manipulation, the strength of the model in fault diagnosis for this test can be rated as exceptional.

## 4. Discussion

Comparison of classification accuracies with different bearing fault detection methods using the same benchmark datasets is shown in Table 6.

**Table 6 - comparison between other methods in the literature and our proposed model**

| Classifier | Data pre-processing | IMS test accuracy | CWRU test accuracy |
|---|---|---|---|
| KNN[9] | HOCs and WT | - | 91.23% |
| SVM[10] | WP | 62.5% | 98.7% (4-classes and small number of samples) |
| SVM ensemble[10] | WP | - | 89.8%-100% (4-classes and small number of Samples) |
| SVM[34] | Statistical locally linear embedding | - | 77.8-94.1% (4-classes and small number of samples) |
| DNN with temporal coherence[35] | - | 94.9% | 94.4% (provide no architecture for their claim) |
| Compact 1D CNN[12] | Filtering-Decimation-Normalization | 93.9% | 93.2% |
| Deep output kernel learning[2] | - | - | - (provided the training accuracy not testing accuracy) |
| Our proposed CRNN | - | 0.9713 | 0.9977 |

In the literature, almost all the previous methods have used some sort of data pre-processing. For instance, Filtering, higher-order-cumulants (HOCs), wavelet transform (WT), wavelet packet transform (WP). Then the best set of features is selected from high dimensional extracted features by applying various dimension reduction techniques such as principal component analysis. For classification of the selected features, various classifiers have been used; although, we can observe that over time, tendencies have shifted from simpler supervised learning models such as SVM and KNN to more complex learning models such as CNN and DNN. The main drawback of almost all these studies is that they have used selected data or manipulated features to increase the accuracy. The manipulated features may not represent the characteristics of the practical system's signal under all circumstances. Consequently, this approach limits the general applicability of those solutions. Moreover, extracting high dimensional features along with necessary post-processing or feature selection methods can significantly increase the cost and computational complexity of the whole system[12]. The next point is that, although many studies have reported an acceptable classification accuracy, their results have generally been limited to relatively small train and test datasets. We used a relatively bigger data-frame of learning features compared to the other articles mentioned in Table 6, and reached a higher accuracy, voiding any data pre-processing or manipulation of features. Moreover, some previous studies claimed to reach a high accuracy although they did not provide their proposed network's architecture and the feasible way through the accuracy [2,35]. In this paper, the network architecture of which was used to achieve the high accuracy, and the experiments behind choosing every single element of the proposed network was presented. Consequently, the advantages of our model are: it is an end-to-end system that has a high resistance against overfitting, can handle big datasets at the shortest possible time, reaches one of the highest accuracies in the literature voiding any sort of pre-processing. In addition, the architecture and step by step procedure are explained clearly which was barely observed in the previous works fulfilled in the field of bearing fault diagnosis.

## 5. Conclusion

In this work, the performance of a generic real-time induction bearing fault diagnosis based on a newly supervised D.L approach was extensively studied. The intelligent system employs a CRNN classifier that is fed by raw time-domain features that are reshaped in the form of tensors of time sequence. Using the end-to-end feature extraction model, the raw bearing vibration data is trained automatically and properly. The model can diagnose the fault precisely and considering the big dataset, in a relatively short time. Implementing the proposed method in a practical situation and industrial-scale has the following advantages compared to the other methods:

- Monitoring a larger and more comprehensive recorded data because the model is highly resistant to overfitting.
- Reaching a more accurate prediction, compared to the other articles in the literature, at a relatively shorter time and the number of epochs.
- The model is end-to-end and can be fed by the raw vibration data directly and no data pre-processing, pre-determined transformation (such as FFT or DWT), manipulated feature extraction and feature selection is required.
- The calculation is more cost-effective compared to some solutions containing data-pre-processing and some complex deep architectures in the literature.

The CRNN classifier-based fault diagnosis system is tested for bearing fault diagnosis using two benchmark vibration datasets. The experimental results validate the effectiveness and feasibility of the CRNN classifier in fault diagnosis. The classifier achieved overall classification accuracies of 97.13% for IMS and 99.77% for CWRU bearing datasets. Classification results demonstrated that the CRNN model can learn highly discriminative features directly from the raw input sensor data.

## 5.1 Future work

Testing and evaluation of a GRU-LSTM network using the collected motor current signal instead of vibration signal as well as reducing the calculation time of the system. Plus, working on unbalanced small scaled datasets using a novel generative adversarial network (GAN) such as the work fulfilled in [36] would be the future work.

### Acknowledgements

The authors are grateful to Dr.Maryam Amirmazlaghani for her helpful comments and constructive suggestions.